# Terahertz superlattice parametric oscillator

K. F. Renk*, B. I. Stahl, A. Rogl, and T. Janzen

*Institut für Angewandte Physik, Universität Regensburg, 93040 Regensburg, Germany*

D. G. Pavel'ev and Yu. I. Koshurinov

*Department of Radiophysics, Nizhny Novgorod State University, Nizhny Novgorod, Russia*

V. Ustinov and A. Zhukov

*A.F. Ioffe Physico-Technical Institute, St. Petersburg, Russia*

**We report a GaAs/AlAs superlattice parametric oscillator. It was pumped by a microwave field (power few mW) and produced 3rd harmonic radiation (frequency near 300 GHz). The nonlinearity of the active superlattice was due to Bragg reflections of conduction electrons at the superlattice planes. A theory of the nonlinearity indicates that parametric oscillation should be possible up to frequencies above 10 THz. The active superlattice may be the object of further studies of predicted extraordinary nonlinearities for THz fields.**

*karl.renk@physik.uni-regensburg.de

A semiconductor superlattice can show a negative differential resistance due to Bragg reflections of the conduction electrons at the superlattice planes [1, 2]. And a superlattice in a negative-differential resistance state can be a gain medium for high-frequency radiation as predicted [3] and concluded from a voltage-induced anomalous THz transmissivity of an array of superlattices [4]. An oscillator with a voltage-biased superlattice as the active element is a candidate for a room-temperature, coherent radiation source in the THz frequency range (see, e.g. Refs. 4 and 5); at present the quantum cascade laser (frequency near 4 THz), operated at liquid nitrogen temperature, is a first compact coherent THz radiation source [6]. In this Letter, we show, experimentally and theoretically, that the nonlinearity mediated by Bragg reflections of the conduction electrons in a superlattice can be used to operate a parametric oscillator suitable for generation of sub-THz and, possibly, THz radiation.

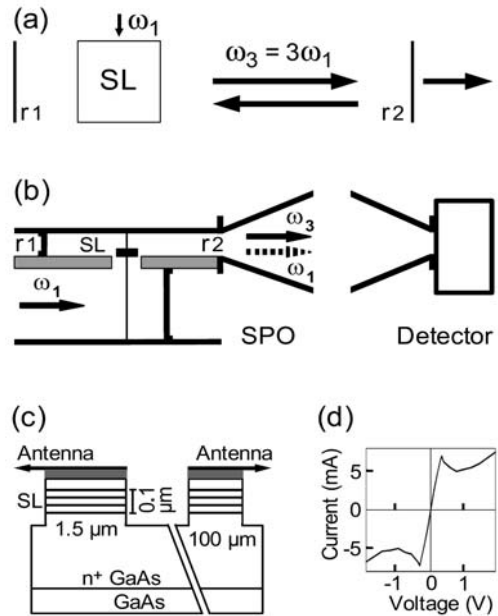

**Fig. 1.** (a) Principle of the semiconductor superlattice parametric oscillator with a superlattice (SL)-active medium. (b) Superlattice parametric oscillator (SPO) and detector (mixer). (c) Superlattice in a quasiplanar design. (d) Experimental current-voltage curve, measured with a static voltage source.

.

The superlattice parametric oscillator (Fig. 1a) is pumped by a microwave field (pump frequency $\omega_1$) and generates 3$^{rd}$ harmonic radiation (frequency $\omega_3$). A resonator (reflector r1, partial reflector r2) for the 3$^{rd}$ harmonic radiation delivers feedback necessary for the initiation and maintenance of parametric oscillation [7, 8]. The superlattice, mounted in a waveguide structure (Fig. 1b), was coupled to a 3$^{rd}$ harmonic-waveguide resonator having a fixed backshort (r1) and with r2 being formed by a mismatch between output port and horn, and to a pump waveguide. Radiation was detected with a thermal detector (Golay cell or powermeter) or a frequency mixer connected to a spectrum analyzer. A 3$^{rd}$ harmonic signal and a pump radiation-leakage signal were measures for the corresponding fields at the site of the superlattice. For pumping we used radiation of a frequency synthesizer (power 4 mW). In a quasiplanar design (Fig. 2c), the superlattice (diameter ~ 1.5 µm; length 0.1 µm; doping $10^{18}$ cm$^{-3}$) was connected to an antenna and, via an n$^+$ GaAs layer and a large-size superlattice (of low resistance) to another antenna. The current-voltage curve (Fig. 2d) showed, above a critical voltage $U_c$ (~ 0.2 V), a negative resistance together with current jumps due to the formation of space-charge domains [9]. The system (Figs. 1b, c, d) has been described in detail elsewhere as a frequency multiplier based on transient domains induced by the microwave pump field [10]. A first sign of parametric oscillation was a threshold behavior of the frequency characteristic of the 3$^{rd}$ harmonic signal $S_3$ (Fig. 2a). At a threshold pump power ($P_{th}$), the oscillation range was narrow. With increasing pump power ($P_1$), the frequency characteristic broadened and extended at large $P_1$ over a range of about 0.3 percent of the centre frequency. Both $S_3$ and the pump-leakage signal $S_1$, measured at the centre frequency, depended strongly on $P_1$ (Fig. 2b). $S_3$ increased, with increasing $P_1$, nonlinearly at small $P_1$ as expected for conventional frequency tripling, showed an abrupt increase at $P_{th}$ and saturated at larger $P_1$. We attribute $P_{th}$ to the threshold of parametric oscillation [11]. The pump-leakage signal $S_1$ increased linearly with $P_1$ up to a critical power ($P_c$) where a strong increase occurred, followed by a weaker increase. We associate $P_c$ with the onset of a negative differential resistance and the corresponding amplitude $\hat{U}_1$ of the pump field with the critical voltage ($\hat{U}_1 \sim U_c$). The strong increase of $S_1$ above $P_c$ indicated an improvement of the matching of the superlattice to the pump waveguide. Within the pump power range from $P_c$ to a power level (slightly above $P_{th}$) which corresponded to the onset of saturation of $S_3$, the signal $S_1$ increased strongly (by a factor of ~ 12) indicating that the amplitude of the pump voltage across the superlattice increased strongly (from $\hat{U}_1 \sim U_c$ to ~ 3.5 $U_c$). We found that a decrease of the reflectivity of the partial reflector (r2) resulted in an increase of $P_{th}$ (up to 10 %) which revealed the occurrence of feedback; strong feedback effects occurred also when we coupled the waveguide resonator to an external resonator for the 3$^{rd}$ harmonic field. The oscillator delivered, at optimum output coupling, a power of about 0.1 mW.

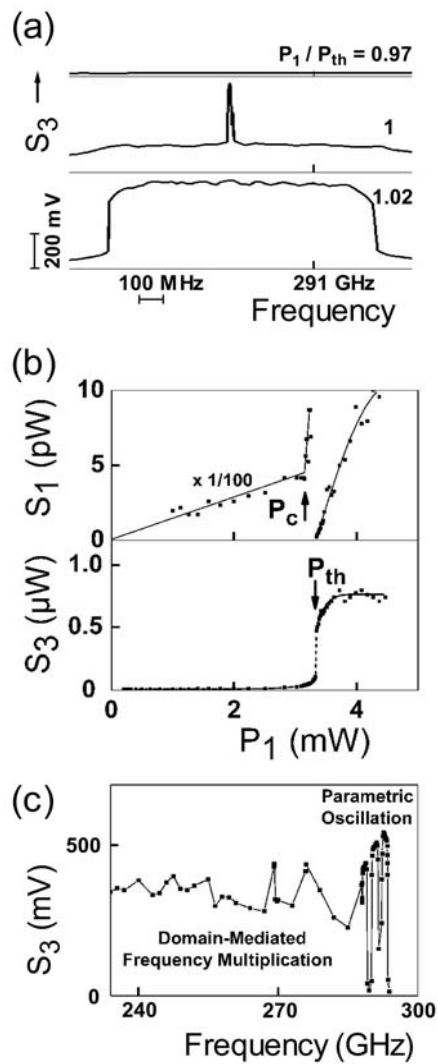

**Fig. 2.** (a) Frequency characteristics of the parametric oscillator. (b) Pump-leakage signal ($S_1$) and 3$^{rd}$ harmonic signal ($S_3$) for different levels of the pump power $P_1$. (c) Domain-mediated frequency multiplication versus parametric oscillation.

We also observed (Fig. 2c) domain-mediated frequency multiplication, distinguished by continuous tuneability over a very wide frequency range, and by its insensibility against feedback by a resonator, and by not having a sharp pump threshold. A simulation of the domain-mediated frequency multiplication showed that the conversion efficiency was also about 10 percent, in agreement with the experimental result. The saturated signal (Fig. 2c) decreased (in the average) slightly towards the higher frequencies and extended up to a frequency (285 GHz; $\omega_3 \tau \sim 0.3$ where $\tau$ is an average relaxation time of our superlattice at room temperature) where parametric oscillation took over; a structure in the frequency characteristic of the parametric oscillation range can be attributed to coupling of the 3rd harmonic field to the pump waveguide.

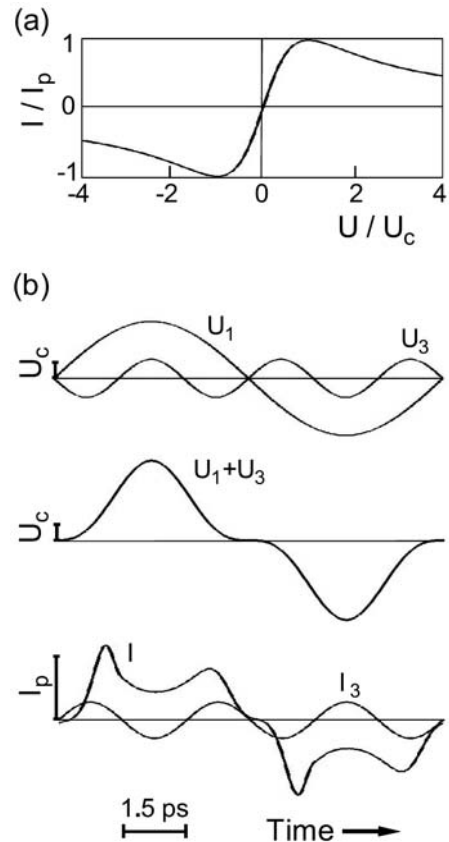

**Fig. 3.** (a) Esaki-Tsu current voltage curve. (b) Voltages and currents involved in the parametric oscillation.

We attribute the parametric oscillation to single-electron dynamics. For an illustration of the gain mechanism, we regard the Esaki-Tsu current-voltage curve [1] (Fig. 3a). The curve is nonlinear, it shows a current peak ($I_p$) at the critical voltage ($U_c$) and a negative differential resistance above $U_c$. The mechanism of parametric gain is sketched in Fig. 1b. A microwave pump voltage ($U_1$) of an amplitude larger than $U_c$ and a 3rd harmonic voltage ($U_3$) add to a total voltage ($U_1 + U_3$). For an instantaneous voltage $U = \hat{U}_1 \cos(\omega_1 t) + \hat{U}_3 \cos(\omega_3 t)$ across the superlattice, optimum gain is expected for the amplitudes $\hat{U}_1 \sim 3U_c$ and $\hat{U}_3 \sim U_c$. Then, ($U_1 + U_3$) is almost zero for a noticeable time and changes rapidly to values above $U_c$, i.e. the superlattice is switched fast between transient states of positive and negative differential resistance.

The current (I) contains a 3$^{rd}$ harmonic current (I$_3$) of opposite phase relative to U$_3$, i.e. the product U$_3$I$_3$ is negative, indicating gain for the 3$^{rd}$ harmonic.

For a theoretical analysis we use the dispersion relation $\varepsilon = \frac{1}{2}\Delta(1-\cos(ka))$ where $\varepsilon$ is the energy and *k* the wave vector for the motion along the superlattice axis, *a* is the superlattice period, and $\Delta$ the width of a miniband formed according to the superlattice periodicity. A Boltzmann transport equation leads to the time dependent drift velocity [12]

$$v(t) = 2v_p \int_{-\infty}^{t} \frac{dt_0}{\tau} \exp\left(-\frac{t-t_0}{\tau}\right) \cdot \sin\left[-\int_{t_0}^{t} \frac{ea}{\hbar} E(t_1) dt_1\right] \quad (1)$$

where E(t) = U(t)/L is the instantaneous field strength, $v_p = \frac{1}{4\hbar}\Delta a\, I_1(x)/I_0(x)$ is the peak drift velocity, $I_1$ and $I_0$ are the first and zeroth order Bessel functions and $x = \Delta/(2k_B T)$, and *L* the superlattice length (*T* temperature, $\hbar$ Planck's constant, *e* elementary charge, $k_B$ Boltzmann constant). The sine term in eq. (1) describes the Bragg reflections [13] and the exponential term the electron relaxation; the different time scales (t$_1$, t$_0$, t) take into account that the relaxation is a statistical process. By Fourier analysis, we determined the amplitudes of the drift velocities $\hat{v}_i$ (i = 1 or 3) and of the current ($\hat{I}_i/I_p = \hat{v}_i/v_p$) using for I$_p$ the experimental value and we calculated the power $P_3 = \frac{1}{2}\hat{I}_3\hat{U}_3$ as well as the resistance $R_3 = \hat{U}_3/\hat{I}_3$ for the 3$^{rd}$ harmonic field.

Results for $\hat{U}_1 = 3U_c$ are shown in Fig. 4a. We find a region of gain (P$_3$ < 0, R$_3$ < 0) on the $\hat{U}_3$ scale, separated by a point of zero conductance (R$_3^{-1}$ = 0) from a range of loss. Optimum gain is found for $\hat{U}_1 = U_c$, i.e. for the Bragg-reflection frequency, $\hat{\omega}_B = \frac{1}{\hbar}ea\hat{U}_3/L$, being equal to the relaxation rate ($\hat{\omega}_B = \tau^{-1}$). In the range of optimum gain, P$_3$ varies only slightly while |R$_3$| shows a strong increase due to the singularity. Accordingly, the active superlattice has the ability to match itself to the 3$^{rd}$ harmonic resonator, via the R$_3(\hat{U}_3)$ dependence and, additionally, to the pump waveguide via the R$_1(\hat{U}_1)$ dependence; the waveguide impedances were between 100 Ω and 200 Ω.

It follows for the case of optimal gain ($\hat{U}_1 = 3U_c; \hat{U}_3 = U_c$) that the power conversion efficiency (~ 0.1) for conversion of pump to 3$^{rd}$ harmonic power is almost independent of $U_c$ and $I_p$ [14].

The calculated 3$^{rd}$ harmonic power is in accordance with the experiment if 3$^{rd}$ harmonic-radiation losses to the pump waveguide and to the superlattice-series

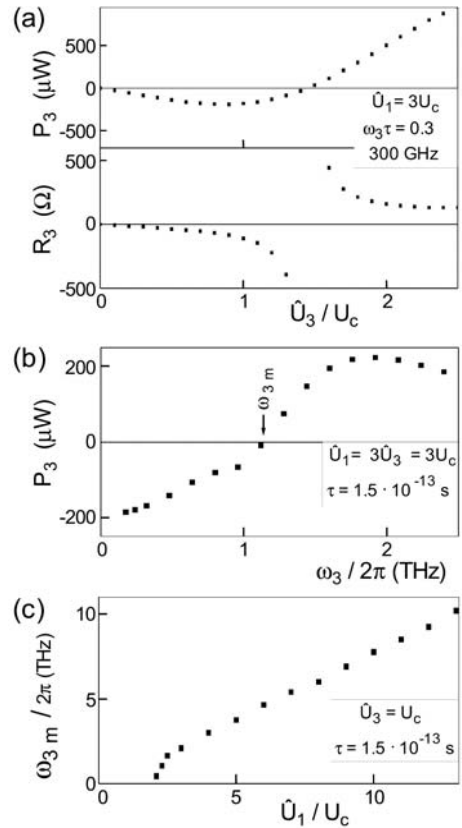

**Fig. 4.** (a) Power $P_3$ and resistance $R_3$ of the 3$^{rd}$ harmonic field for different amplitudes of the 3$^{rd}$ harmonic field. (b) 3$^{rd}$ harmonic power $P_3$ for different frequencies. (c) Maximum oscillation frequency $\omega_{3m}$ for different pump-voltage amplitudes ($\hat{U}_1$).

resistance are taken into account. A power balance shows that a 3$^{rd}$ harmonic photon made in the average about three passages through the active superlattice, delivering the gain coefficient ($3 \cdot 10^4$ cm$^{-1}$) and the gain-cross section ($3 \cdot 10^{-14}$ cm$^2$) for an electron, in accordance with the calculated spatial amplitude (~ 1 nm) of the drift oscillation of an electron at the frequency $\omega_3$. The frequency dependence (Fig. 4b) indicates that |$P_3$| decreases, in the range of gain, with increasing $\omega_3$ up to a

maximum frequency $\omega_{3m}/2\pi$ (~ 1 THz). The maximum frequency (Fig. 4c) can be raised to the THz range if besides the pump frequency also the pump-voltage amplitude is increased; $\Delta$ has to be appropriately chosen to fulfil the condition $\hbar\omega_3 < \Delta$. Parametric oscillators of higher power in series or, alternatively, a parametric oscillator working on a higher than the 3$^{rd}$ harmonic, should allow reaching, with a conventional microwave pump source, the THz range.

Domain-mediated tripling has a frequency cut-off near 1 THz (where $\omega_3\tau \sim 1$ [15]). Our experiment indicates that, for a microwave-pumped superlattice, a transient-domain state can be overcome already in the sub-THz range ($\omega_3\tau \sim 0.3$) by the single-electron nonequilibrium gain state [16, 17].

In conclusion, we demonstrated the operation of a parametric oscillator with a superlattice as the active medium having a high gain that is based on Bragg reflections of the conduction electrons, and presented a theoretical analysis suggesting that a superlattice parametric oscillator should be achievable at THz frequencies. We also showed that under appropriate experimental conditions a domain state can be avoided in favour of a single-electron nonequilibrium state, which can be the basis for further experimental studies of this state with extraordinary nonlinear properties at sub-THz and THz frequencies. Due to attractive properties, namely narrow bandwidth, coherence, compactness, room-temperature operation, and the wide tuneability, the superlattice parametric oscillator appears to be a promising optoelectronic device.

**Acknowledgement**

The experimental realization of the parametric oscillation was first recognized by K.F.R. Experiment and theory were performed in Regensburg, construction of the frequency multiplier and superlattice structuring as well as a simulation of the domain-mediated frequency multiplication in Nizhny Novgorod, and the preparation of the superlattice in Sankt Petersburg. One of us (K.F.R.) would like to thank L. Keldysh for fruitful discussions. The work has been supported by the Deutsche Forschungsgemeinschaft and the Russian Foundation for Basic Research.